\documentstyle[epsf,referee]{l-aa-mpa}
\newcommand{\be}{\begin{displaymath}}
\newcommand{\ee}{\end{displaymath}}
\newcommand{\omcen}{$\omega$~Cen}
\newcommand{\delmix}{$\delta M_{\mbox{\tiny mix}}$}
\newcommand{\dmix}{$D_{\mbox{\tiny mix}}$}
\newcommand{\cc}{\mbox{$^{12}$C/$^{13}$C}}
\newcommand{\NeNa}{$^{22}$Ne(p,$\gamma$)$^{23}$Na}

\begin{document}

\thesaurus{06(08.01.1; 08.05.3; 08.09.3; 10.07.2)}

\title{The puzzling MgAl anticorrelation in globular-cluster red giants:
primordial plus deep mixing scenario?}

\author{P.A.~Denissenkov\inst{1,2,3}, G.S.~Da~Costa\inst{1},
J.E.~Norris\inst{1}, A.~Weiss\inst{3}}

\institute{Mount Stromlo and Siding Spring Observatories, Institute of
           Advanced Studies, The Australian National University,
           Private Bag, Weston Creek P.O., ACT 2611, Australia
	   \and
	   Astronomical Institute of the St. Petersburg University,
	   Bibliotechnaja Pl.~2, Petrodvorets, 198904~St.\,Petersburg,
	   Russia
           \and
           Max-Planck-Institut f\"{u}r Astrophysik,
	   Karl-Schwarzschild-Str.~1, 85740 Garching,
	   Federal Republic of Germany}

\offprints{A.~Weiss (weiss@mpa-garching.mpg.de)}

\date{Received; accepted}

\maketitle
\markboth{P.A.~Denissenkov et al.:
The puzzling MgAl anticorrelation: primordial plus deep mixing scenario?}
{P.A.~Denissenkov et al.:
The puzzling MgAl anticorrelation: primordial plus deep mixing scenario?}

\clearpage
\thispagestyle{empty}
\mbox{~}
\vfill
\begin{abstract}

   Star-to-star abundance variations of C, N, O, Na and Al in
globular-cluster red giants have been recently supplemented by the finding
that [Mg/Fe] is depleted in stars with extremely large [Al/Fe] (Shetrone
1996a).  To find out which of the magnesium isotopes is responsible for the
observed depletion of [Mg/Fe] Shetrone (1996b) also undertook an isotopic
analysis of Mg and found that it is $^{24}$Mg which is depleted in Al-rich
giants. On the other hand, Norris \& Da Costa (1995) demonstrated that even
in the massive globular cluster \omcen\ which has intrinsic spreads in both
[Fe/H] and the abundances of the s-process elements, [O/Fe] anticorrelates
with [Na/Fe] and [Al/Fe] as in ``normal'' monometallic clusters. These new
spectroscopic results allow us to test current models of stellar evolution
and nucleosynthesis, as well as those of the formation and chemical
enrichment of globular clusters. In an effort to explain self-consistently
these observations we have considered two possibilities: (1) a deep mixing
scenario which assumes that in red giants some kind of (extra)mixing
transports products of nuclear reactions from the hydrogen burning shell
(HBS) to the base of the convective envelope; and (2) a combination of
primordial and deep mixing scenarios. It is shown that (1) cannot account for
the anticorrelation of [Mg/Fe] vs. [Al/Fe] without additional {\it ad hoc}
assumptions, among which we identify a strong but still undetected low energy
resonance in the reaction $^{24}$Mg(p,$\gamma)^{25}$Al, and episodical
increases of the HBS temperature up to the value $T\approx 74\,10^6$\,K. In
(2) intermediate mass AGB stars are assumed to produce the decreased
$^{24}$Mg and increased $^{25}$Mg initial abundances in some globular-cluster
low mass stars and Al is synthesized at the expense of $^{25}$Mg in the HBS
and transported to the surface of the red giant by extramixing. We discuss
advantages and deficiencies of both scenarios and propose some observational
tests.


\end{abstract}
\vfill
\mbox{~}

%==================== Introduction ====================================
\section{Introduction}

   Recently Shetrone (1996b, hereafter S96) has determined magnesium isotopic
compositions for a small sample of bright red giants in the globular cluster
M~13.  His finding that many giants in this system have
[($^{25}$Mg+$^{26}$Mg)/$^{24}$Mg] $\sim$ +0.4, in stark contrast to values
$\sim$ --0.4 found in field halo stars (see also McWilliam \& Lambert 1988
and references therein), is of fundamental importance, and highlights yet
again the differences in abundance patterns found in cluster and field stars.
Shetrone's work followed the earlier discovery that large Al enhancements
were usually accompanied by moderate Mg depletions in this cluster (Shetrone
1996a), as well the demonstration that anticorrelations of Na and Al with O
exist not only in ``normal'' monometallic clusters such as M~13 (see Kraft et
al. 1997) but also in the massive globular cluster \omcen\ whose stars show
intrinsic spreads in both metallicity and the abundances of the s-process
elements (Norris \& Da~Costa (1995, hereafter ND95)).  These new data
challenge stellar evolution theory to provide an appropriate explanation of
the abundance anomalies, on the one hand, and present a number of new
observational constraints on possible primordial processes responsible for
abundance variations in globular clusters, on the other. In this paper we use
the latest spectroscopic results to address the question of whether the
modern theory of nucleosynthesis in stars can self-consistently reproduce the
whole spectrum of abundance variations seen in globular-cluster red giants
(GCRGs) and, if not, to seek constraints on additional {\it ad hoc}
assumptions which make such reproduction possible.

   Stars leaving the main sequence (MS) in present-day globular clusters have
small masses $M\approx 0.8 - 0.9\,M_\odot$ and very low metallicities,
covering the range --2.4 $<$ [Fe/H] $<$ --0.2, corresponding to $8\,10^{-5}$
$<$ Z $<$ 0.01, (where we adopt [Fe/H] = $lg(Z/Z_\odot$), and
$Z_\odot=0.01886$ (Anders \& Grevesse 1989))\footnote{We use the standard
spectroscopic notation, i.e. [A/B]$=\lg(N(\mbox{\rm A})/N(\mbox{\rm
B}))_{\mbox{\tiny star}}-\lg(N(\mbox{\rm A})/N(\mbox{\rm B}))_\odot$, where
$N$(A) and $N$(B) are atomic number densities of the nuclides A and B,
respectively. Z is the mass fraction of elements heavier than
helium.}. Standard stellar evolution theory has no doubts about subsequent
structural and chemical histories of such stars, at least until the beginning
of the core helium flash. MS central hydrogen burning, which was dominated by
pp-chains, is now replaced by shell hydrogen burning with the main energy
output provided by the CNO-cycle. The shell advancing outwards in mass causes
a gradual growth of the underlying helium core until the core mass becomes
large enough to trigger the helium flash. These internal structural changes
are accompanied by the star ascending the red giant branch (RGB). Its surface
chemical composition is not expected to change significantly during this
evolutionary stage. The only important event between the MS turn-off and the
core helium flash is the well-known first dredge-up episode which takes place
on the subgiant branch. The displacement of the star from the MS to a cooler
region of the HR diagram favours the convective envelope extending its base
into deep layers which were in radiative equilibrium on the MS and where some
mild transformations among CN isotopes in the then energetically unimportant
CN-cycle occurred.  As a result the surface abundances of $^{12}$C, $^{13}$C
and $^{14}$N (as well as those of $^7$Li, $^3$He and a few other light
nuclides which are not discussed in this work; for recent theoretical results
on their standard dredge-up and non-standard deep mixing evolutionary changes
see Sackmann \& Boothroyd (1997), Weiss et al. (1996) and Charbonnel (1995))
begin to decline from their initial values. This excursion of the base of the
convective envelope (BCE) into the interior of the star continues until it
arrives at its deepest penetration, which corresponds to the end of the first
dredge-up; after that, convection begins to retreat and later follows
approximately the outward advancement of the hydrogen burning shell
(HBS). Our standard evolutionary calculations for a $M=0.8\,M_\odot$ star
show that, depending on Z, the first dredge-up changes of the surface
$^{12}$C, $^{13}$C and $^{14}$N abundances are indeed very modest: $(Z,
^{12}\mbox{\rm C}/^{13}\mbox{\rm C},\Delta\lg ^{12}\mbox{\rm C}, \Delta\lg
^{14}\mbox{\rm N}$) = (10$^{-4}$,\,64,\,\mbox{--0.0024},\,0.0025),
(5\,10$^{-4}$,\,50,\,\mbox{--0.0064},\,0.012),
(5\,10$^{-3}$,\,45,\,\mbox{--0.0092},\,0.021) (on the MS \cc$\approx 90$;
here and in what follows a nuclide's chemical symbol is used also to denote
its number density).

   When comparing these predictions of standard stellar evolution theory with
available observational data on the atmospheric chemical composition of red
giants in globular clusters one finds obvious and numerous
disagreements. {\it (i)} The measured $^{12}$C/$^{13}$C ratios have very low
values often approaching the limit 3.5 which is a characteristic of the
equilibrium CN-cycle (Smith \& Suntzeff 1989; Brown \& Wallerstein 1989;
Suntzeff \& Smith 1991; Brown, Wallerstein \& Oke 1991; Briley et al. 1994;
S96). {\it (ii)} The differences in $\lg$C and $\lg$N among red giants in the
same cluster reach an order of magnitude or even more. Moreover, carbon
abundances [C/Fe] are found to anticorrelate with luminosity (to correlate
with the absolute visual magnitude), consistent with a progressive C
depletion as the star ascends the RGB at luminosities well above the end of
the canonical first dredge-up (as seen, for example, in M~92 (Langer et
al. 1986), NGC~6397 (Briley et al. 1990), M4 and NGC~6752 (Suntzeff \& Smith
1991)). {\it (iii)} The most intriguing result is that in many globular
clusters there are large (up to 1~dex) variations of O and Na among red
giants (Kraft 1994) and in some clusters Al and Mg abundances also vary from
star to star (Kraft et al. 1997).

   It is important to note that all nuclides mentioned above whose abundances
show considerable scatter in GCRGs ($^{12}$C, $^{13}$C, N, O, Na, Mg isotopes
and Al) are potential participants in hydrostatic hydrogen burning. During
transformation of H into He in the CNO-cycle, which plays the chief role in
this process, the relative abundances of the CNO nuclides are changing
whereas their net sum remains constant, i.e. the CNO nuclides actually work
as catalysts. So do the NeNa and MgAl nuclides in the NeNa- and MgAl-cycles,
respectively.  Detailed flow-charts and the latest estimates of reaction
rates for all three cycles can be found in the review of Arnould et
al. (1995). Such approximate constancy of the sum C+N+O, despite the large
spreads in the individual abundances of C, N and O, has been observed in
M\,13 and M\,3 by Smith et al. (1996), in NGC\,362 and NGC\,288 by Dickens et
al. (1991) and in \omcen\ by ND95. There is also some evidence of constancy
of the sum Mg+Al in M~13 (Kraft et al. 1997).

   In contradistinction, the heavier $\alpha$ elements such as Si, Ca and Ti,
which are believed to be produced by successive $\alpha$-captures in massive
stars and which do not participate in hydrostatic hydrogen burning, do not
show any scatter in GCRGs.  Instead their mean abundances agree well with the
abundances of $\alpha$-elements observed in field Population~II dwarfs,
i.e. $\langle[\alpha$/Fe$]\rangle\approx+0.4$. In addition, the iron peak
elements Cr and Ni, synthesized during SNe explosions, do not exhibit
abundance anomalies. As regards Fe itself, in most globular clusters its
abundance is surprisingly constant within the same cluster. The only
exception of the rule is \omcen\ (M~22 may be another) where the abundance
range $-1.8<$[Fe/H]$<-0.8$ and the rise of abundances of s-process elements
with that of Fe are certainly records of the cluster's more complicated
chemical enrichment history (ND95).

   To summarize, observations strongly support the idea that the star-to-star
abundance variations (or, more precisely, at least a substantial part of
them) in globular clusters were presumably produced during hydrogen
burning. The fact that there are fairly good correlations between
overabundances of N, Na and Al, on the one hand, and underabundances of C, O
and Mg, on the other hand, is undoubtedly a sign of these abundance
anomalies' simultaneous origin. The next question to answer is the place
where these variations were produced. In our attempt to solve this puzzle we
have chosen the latest spectroscopic data on the chemical composition of red
giants in the globular clusters M~13 (Kraft et al. 1997; S96) and \omcen\
(ND95). The former cluster is an extreme representative of ``normal''
(i.e. showing neither iron nor neutron-rich elements abundance variations)
which contributes much to the ``global anticorrelation'' of [O/Fe] versus
[Na/Fe] (see the review paper by Kraft 1994). \omcen, on the other hand,
provides evidence that the O vs. Na and O vs. Al anticorrelations are not a
prerogative of ``normal'' clusters.  Furthermore, \omcen\ is the only
globular cluster not obeying the ``global anticorrelation'' since it contains
giants with extremely high [Na/Fe] (up to 1~dex) well above the general trend
seen in ``normal'' clusters where [Na/Fe] saturates at $\sim$ 0.5~dex.  We
shall see how theory can interpret this anomaly.

   The remainder of the paper is organized as follows. In Sect.~2 we briefly
describe our theoretical tools -- computer codes used to perform stellar
evolution and nucleosynthesis calculations. Section~3 deals with the most
promising candidate for the origin of the star-to-star abundance variations
in globular clusters -- deep mixing in evolving red giants. This, however,
provides an incomplete solution to the problem.  Sources of additional
primordial pollution (SNe explosions, AGB stars and a ``black box'') which
can contribute and modify the deep mixing scenario are discussed in Sect.~4.
(We refer the reader also to Smith and Kraft (1996) who suggest that Ne novae
may provide the additional pollution.  In the present work we shall not
address this possibility.)  Our concluding remarks are given in Sect.~5.
   
%======================================================================

%==================== Computer codes used in the paper ================
\section{Computer codes used in the paper}

   To calculate the evolution of a $M=0.8\,M_\odot$ star from the ZAMS
through to the core helium flash we used the code described by Raffelt
\& Weiss (1992), which, in the meantime, has been updated with respect
to new opacity data (Rogers \& Iglesias 1992; Iglesias \& Rogers 1996)
and plasma neutrino emission (Haft, Raffelt \& Weiss 1994).
The mass fraction of heavy elements was taken as $Z=5\,10^{-4}$ ($\lg
Z/Z_\odot =-1.58$) to approximately match the metallicities of M~13 ([Fe/H] =
--1.49) and \omcen\ ($-1.8<$[Fe/H]$<-0.8$), whereas the initial helium
abundance was $Y=0.24$ as in the big bang composition (Walker et al. 1991).
The adopted value of the mixing length parameter $\alpha =1.48$ (times the
pressure scale height, $H_P$) came from calibrating a solar model.

   A number of selected red giant models were used to interpolate temperature
and density distributions as well as to follow the movements in mass of the
HBS and the BCE as the star ascends the RGB. Following the procedures
described in Denissenkov \& Weiss 1996 (Paper~I), mixing in the radiative
zone between the HBS and the BCE was modelled by introducing diffusion terms
into the standard equations of nuclear kinetics.  Results of these deep
mixing calculations depend on a choice of two parameters: the depth of mixing
\delmix\ determined as a relative mass coordinate of the deepest radiative
layer involved in the mixing (measured from the HBS in units of the mass
separating the HBS and the BCE), and the diffusion coefficient \dmix. The
latter may not be chosen arbitrarily large because there are some estimates
of its ``reasonable'' upper limit values which yield \dmix$\sim 10^7 -
10^9$\,cm$^2$\,s$^{-1}$ (see Paper I). The nuclear kinetics network used in
the deep mixing calculations includes all important nuclides participating in
the reactions of the CNO-, NeNa- and MgAl-cycles as well as those in a few
reactions of the pp-chains (to follow changes of the $^3$He abundance) and
numbers 26 particles coupled by 30 reactions.

   To estimate nucleosynthesis yields from intermediate mass AGB stars we
employed an algorithm developed by Denissenkov et al. (1997, hereafter
Paper~II) which is very similar to the scheme used by Renzini \& Voli
(1981). The algorithm takes into account nuclear processing in the HBS (in an
AGB star!) and at the BCE (hot bottom burning, hereafter HBB) between pulses
as well as the convective shell He burning nucleosynthesis during a pulse in a
simplified parameterized manner. A number of parameters are not well
constrained in these calculations.  For instance, the temperature of the HBB
and the amount of material dredged up after finishing every pulse remain very
uncertain (Lattanzio et al. 1997). Therefore, in Sect.~4 we shall concentrate
only on those results of our nucleosynthesis calculations in intermediate
mass AGB stars which weakly depend on the choice of such parameters or,
otherwise, specially emphasize which values of parameters lead to a
particular result supported by observations. Unfortunately, mainly because of
uncertainties in the current treatment of convective overshoot in stars
(Frost \& Lattanzio 1996), such an approach seems to be the only one possible
at present. Our nuclear kinetics network applied in AGB stars includes the
same nuclides and reactions as those in the deep mixing code plus Si
isotopes; 3$\alpha$-reaction; $(\alpha,\gamma)$-reactions on $^{12}$C,
$^{14}$C, $^{14}$N, $^{15}$N, $^{16}$O, $^{17}$O, $^{18}$O, $^{20}$Ne,
$^{21}$Ne, $^{22}$Ne, $^{24}$Mg, $^{25}$Mg and $^{26}$Mg;
$(\alpha$,n)-reactions on $^{13}$C, $^{17}$O, $^{18}$O, $^{21}$Ne, $^{22}$Ne,
$^{25}$Mg and $^{26}$Mg; neutron captures by $^{12}$C, $^{13}$C, $^{16}$O,
$^{19}$F, $^{20}$Ne, $^{21}$Ne, $^{22}$Ne, $^{23}$Na, $^{24}$Mg, $^{25}$Mg,
$^{26}$Mg, $^{27}$Al, $^{28}$Si, $^{29}$Si, $^{30}$Si and by an averaged
neutron-sink heavy ``nucleus'' $^{31}$X$_{14}$; as well as the reactions
$^{14}$C(p,$\gamma)^{15}$N, $^{14}$N(n,p)$^{14}$C,
$^{19}$F($\alpha$,p)$^{22}$Ne, $^{26}$Al$^{\mbox{\tiny g}}$(n,p)$^{26}$Mg,
$^{21}$Ne(n,$\alpha)^{18}$O and $^{26}$Al$^{\mbox{\tiny
g}}$(n,$\alpha)^{23}$Na. The total numbers of nuclides and reactions in this
network are 26 and 69, respectively.

   To check the degree to which the s-process nucleosynthesis can contribute
to the yields of heavy nuclei from intermediate mass AGB stars at low
metallicity we have prepared a code which allows all the reactions mentioned
above plus neutron captures by numerous heavier nuclides and solves a nuclear
kinetics network for the total of 409 particles coupled by 1273
reactions. Neutrons are assumed to have their local equilibrium abundances in
every mesh point within the He convective shell.

   In all nucleosynthesis codes we used the same input data for the reaction
rates. For reactions between charged particles the tables of Caughlan \&
Fowler (1988, hereafter CF88) were usually used. Exceptions are the reactions
$^{17}$O(p,$\alpha)^{14}$N and $^{17}$O(p,$\gamma)^{18}$F for which the rates
proposed by Landr\'{e} et al. (1990) with the uncertainty factors $f_1=0.2$
and $f_2=0.1$ recommended by Boothroyd et al. (1995) were preferred, and the
reaction $^{12}$C$(\alpha,\gamma)^{16}$O whose CF88 rate was multiplied by
a constant 1.7 as suggested by Woosley \& Weaver (1995). For the NeNa-cycle
we also considered the latest nuclear reaction rates advocated by El~Eid \&
Champagne (1995) but did not include them in our main calculations (see
Sect.~3). Below, where these new rates for the NeNa-cycle are used instead of
those of CF88 we note this explicitly.

   Neutron capture cross sections were taken from summaries published by
Fowler et al. (1967), Holmes et al. (1976), Woosley et al. (1978), Bao \&
K\"{a}ppeler (1987), Ratynski \& K\"{a}ppeler (1988), Beer, K\"{a}ppeler \&
Arcoragi (1989), Cowan et al. (1991) and Schatz et al. (1995). Beta-decay
rates were interpolated in temperature and density using the tables of
Takahashi \& Yokoi (1987).

   Finally, to calculate the structure of a low metallicity massive ZAMS star
(Sect.~4) we employed a modified version of Paczy\'{n}ski's code (1970)
described by Denissenkov (1990).

   Thus, in total, we made use of five different stellar
evolution/nucleosynthesis codes in this work.

%======================================================================

%==================== Fig.1 ===========================================
\begin{center}
\begin{figure}
\epsfxsize=10.1cm
\hspace{3.0cm}\epsffile{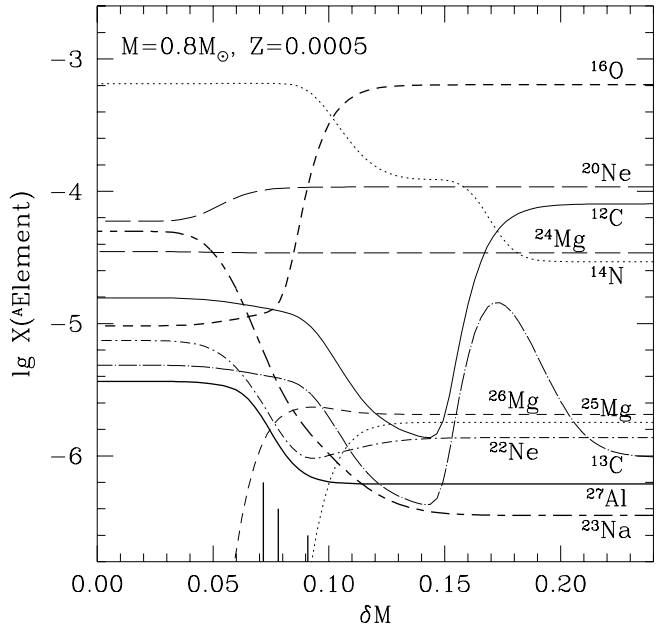}

\caption[]{Abundance profiles for a number of nuclides participating in the
CNO-, NeNa- and MgAl-cycles in radiative layers adjacent to the HBS in a
$M=0.8\,M_\odot$ model star having $\lg(L/L_{\odot})=3.0$ and $Z=5\,10^{-4}$
approximately matching metallicities of the globular clusters M~13 and
\omcen. The mass coordinate $\delta M$ is measured from the HBS in units of
the mass separating the HBS and BCE. The vertical segments on the abscissa
show locations of layers where (from right to left) 1, 5 and 10 percent of H
were consumed
          }

\end{figure}
\end{center}

%======================================================================

%==================== The deep mixing scenario ========================
\section{The deep mixing scenario}

   By the end of the 1970s the fact of star-to-star abundance variations in
GCRGs was quite well established for C, N and to a lesser degree for O. In
1979 Sweigart \& Mengel found that in low metallicity red giant models the
radiative layer where C was transformed into N, and for very low metallicity
even the layer where O was transformed into N, were rather well separated
from the main part of the HBS where H was transformed into He. This meant
that rotationally-driven meridional circulation currents, if present beneath
the BCE, could freely penetrate close enough to the HBS and transport
outwards material with depleted C and O and enhanced N abundances; usually, a
large mean molecular weight gradient forms a barrier which cannot be
penetrated by meridional circulation. Further developments of Sweigart \&
Mengel's idea have been commonly referred to as ``the deep mixing (or
evolutionary) scenario''. It should be noted that in this model the nature of
the mixing mechanism is usually not specified. Exceptions are the pioneering
work of Sweigart \& Mengel (1979) itself and that of Smith \& Tout (1992)
where meridional circulation in its simplest classical treatment was shown
capable of providing the required rate of mixing, and the recent paper of
Charbonnel (1995) who considered the more complicated mixing algorithm
elaborated by Zahn (1992) which takes into account the interaction between
meridional circulation and turbulent diffusion. Other works concentrate on
nucleosynthesis aspects of the problem and try to answer the question of
whether {\it any} postulated mixing can explain the whole spectrum of
abundance variations (and correlations) seen in globular clusters (as does
the present work).

   In the 1980s, following the first reports by Cottrell \& Da Costa (1981)
and Norris et al. (1981) that in NGC 6752 the N enhancements were accompanied
by overabundances of Na and Al, evidence accumulated that this was a common
feature of many clusters.  Moreover, Paltoglou \& Norris (1989) found that in
\omcen\ there is an anticorrelation between Na and O, which anticipated the
discovery of the tight global anticorrelation of [O/Fe] versus [Na/Fe] by
Kraft et al. (1993). Because it was then absolutely unclear how Na and Al,
with rather large nuclear charges (and therefore rather strong Coulomb
barriers against charged particle nuclear reactions), could be produced
during hydrogen burning in low mass stars, the deep mixing scenario had begun
to lose its supporters.

   Its status was rehabilitated by low energy resonances in the reactions
$^{22}$Ne(p,$\gamma)^{23}$Na, $^{25}$Mg(p,$\gamma)^{26}$Al and
$^{26}$Mg(p,$\gamma)^{27}$Al. First, Denissenkov \& Denissenkova (1990)
demonstrated that the temperature in the O depleted layer is high enough for
the reaction $^{22}$Ne(p,$\gamma)^{23}$Na to proceed (due to a resonance!)
even faster than the reaction $^{16}$O(p,$\gamma)^{17}$F responsible for the
O depletion.  Second, Langer et al. (1993) found that $^{27}$Al can also be
synthesized (mainly at the expense of $^{25}$Mg) beneath the O layer in low
metallicity red giants. These findings are illustrated in Fig.~1, where
abundance profiles for a number of nuclides participating in the \mbox{CNO-,}
NeNa- and MgAl-cycles are plotted in the radiative layers adjacent to the HBS
in a $M=0.8\,M_\odot$ model star having $\lg(L/L_{\odot})=3.0$ and
$Z=5\,10^{-4}$. The vertical segments on the abscissa show the locations of
layers where (from right to left) 1, 5 and 10 percent of H have been
consumed. We will not consider mixing penetrating too deeply into the HBS
because: {\it (i)} on theoretical grounds the molecular weight gradient is
expected to stabilize processes inducing both meridional circulation and
turbulent diffusion (e.g. Kippenhahn 1974; Talon \& Zahn 1997); {\it (ii)}
observed Na abundances constrain the mixing depth (and rate) too (see below);
and {\it (iii)} from the computational point of view, deep mixing is not
allowed to bring too much fresh hydrogen fuel into the HBS because otherwise
one would then have to take into account feedback of the mixing on the
internal structure and evolution of red giants.

  For completeness we also refer the reader to the work of Cavallo, Sweigart
\& Bell (1996), and Langer, Hoffman \& Zaidins (1997) who have recently
considered the role of deep mixing in producing the abundance patterns under
discussion here.

\subsection{The initial chemical composition}

   To calculate the abundance profiles in Fig.~1 we adopted the following
initial chemical composition for elements heavier than He: first, all the
relevant solar abundances (Anders \& Grevesse 1989) were multiplied by a
factor $Z/Z_\odot$; then the scaled abundances of $\alpha$-elements
($^{16}$O, $^{20}$Ne, $^{24}$Mg and $^{28}$Si) were increased by a factor
2.512 to agree with the average value
$\langle[\alpha$/Fe$]\rangle\approx+0.4$ inferred from field Population~II
dwarfs (Wheeler et al. 1989); next, abundances of Na and Al were reduced to
[Na/Fe] = [Al/Fe] = --0.4 to take into account the even/odd effect (ibid.);
and finally, to comply with the global anticorrelation of [O/Fe] versus
[Na/Fe] the $^{22}$Ne abundance was also decreased to ensure that
[$^{22}$Ne/Na] = 0 (Paper~I).

   We did not introduce in the initial chemical composition any corrections
resulting from the first dredge-up because they were negligible (Sect.~1)
compared with the surface abundance changes produced by the following deep
mixing.  For completeness we note that in our calculations we adopt isotopic
ratios $^{24}$Mg/$^{25}$Mg/$^{26}$Mg\,=\,90/4.5/5.0, except as otherwise
stated.

\subsection{The CNO and Na abundances}

   Three further details evident in Fig.~1 are important for the following
discussion, and while well-known, are worthy of note. {\it (i)} As one
approaches the HBS the Al abundance increases to a considerably smaller
degree than that of Na, with the latter experiencing two successive rises,
the first of which appears at the expense of $^{22}$Ne while the second is
due to consumption of the much more abundant $^{20}$Ne. {\it (ii)} The
$^{24}$Mg abundance shows no changes at all. It should be emphasized that if
one took a pre-core-helium-flash model star one would still find very little
$^{24}$Mg depletion even well inside the HBS. {\it (iii)} $^{12}$C first
decreases with depth and then, after the CN-cycle comes to equilibrium, goes
up because in CN-cycle equilibrium the abundance of $^{12}$C (and $^{13}$C)
follow that of N. The latter begins to increase further with depth at the
expense of O. Thus the surface abundances of $^{12}$C and $^{13}$C are
expected to behave qualitatively differently from that of, say, O with
changing mixing depth (and rate).  For example, in a star with {\it deeper}
mixing (\delmix$\approx 0.06$) {\it more} C can survive at the surface than
in a star with shallow mixing (\delmix$\approx 0.13$).

   We have found that deep mixing with the parameters of depth and rate
\delmix = 0.05, \dmix = $5\,10^8$\,cm$^2$\,s$^{-1}$ reproduces quite well all
of the observed correlations (or, more precisely, abundance trends) except
that of O vs. Al seen in \omcen, as is shown by the solid lines in Fig.~2.
The observational data in the figure have been taken from ND95. We have
corrected their N abundances (the average shift applied to [N/Fe] was
+0.5\,dex) to agree better with the data of Brown \& Wallerstein (1993) and
Norris \& Da Costa (1997). 

   To explain the anticorrelation of O vs. Na in M\,13, which is a very good
representative of the global anticorrelation of [O/Fe] versus [Na/Fe] in
``normal'' clusters, we needed to assume a somewhat less deep but faster
mixing than in the case of \omcen: \delmix = 0.06, \dmix =
$2.5\,10^9$\,cm$^2$\,s$^{-1}$.  The model is shown by the long-dashed lines
in Fig.~3 where the observed abundances of Na and Al have been taken from
Kraft et al. (1997), and corrected by +0.05\,dex and --0.25\,dex,
respectively, to compensate for differences between the {\it gf} values
adopted in the two works\footnote{These corrections represent the mean
differences between the {\it gf} values of the transitions in common between
the two investigations.  We note here for the record that we also tested for
possible systematic abundances differences which might result from the
different formalisms of the Lick group and ND95.  We found that when we
processed the equivalent widths, {\it gf} values, and atmospheric parameters
of the Lick workers through the ND95 formalism we obtained Na and Al
abundances which agreed with theirs to within $\sim$ 0.05 dex.}.  Note that
here too there is no agreement between the observed and model results for
[Al/Fe] versus [O/Fe]. It is worth commenting that even with the mixing
penetrating as deep as \delmix $\approx$ 0.05 -- 0.06 (cf.\ Fig.\ 1), the
diffusive mixing rate is such that the amount of additional hydrogen
processed in the HBS is relatively small.  In the two cases reported here,
the final surface H abundances are decreased by only 9.9 and 7.5\%,
respectively, and Sweigart (1997) has shown that under these circumstances
the giant branch evolution is essentially unaltered.

   For comparison, the deep mixing solution giving the best fit for the global
anticorrelation is also plotted in Fig.~2 (long-dashed lines).  From Fig.~2a
we infer that \omcen\ may contain red giants exhibiting in their atmospheres
Na produced not only at the expense of $^{22}$Ne but also from $^{20}$Ne.
That is to say, in some \omcen\ stars deep mixing may penetrate even to the
second rise of Na seen in Fig.~1.  To our knowledge this is the only example
of this phenomenon observed to date.

%==================== Fig.2 ===========================================
\begin{figure*}[ht]
\epsfxsize=16cm
\epsffile [80 240 580 680] {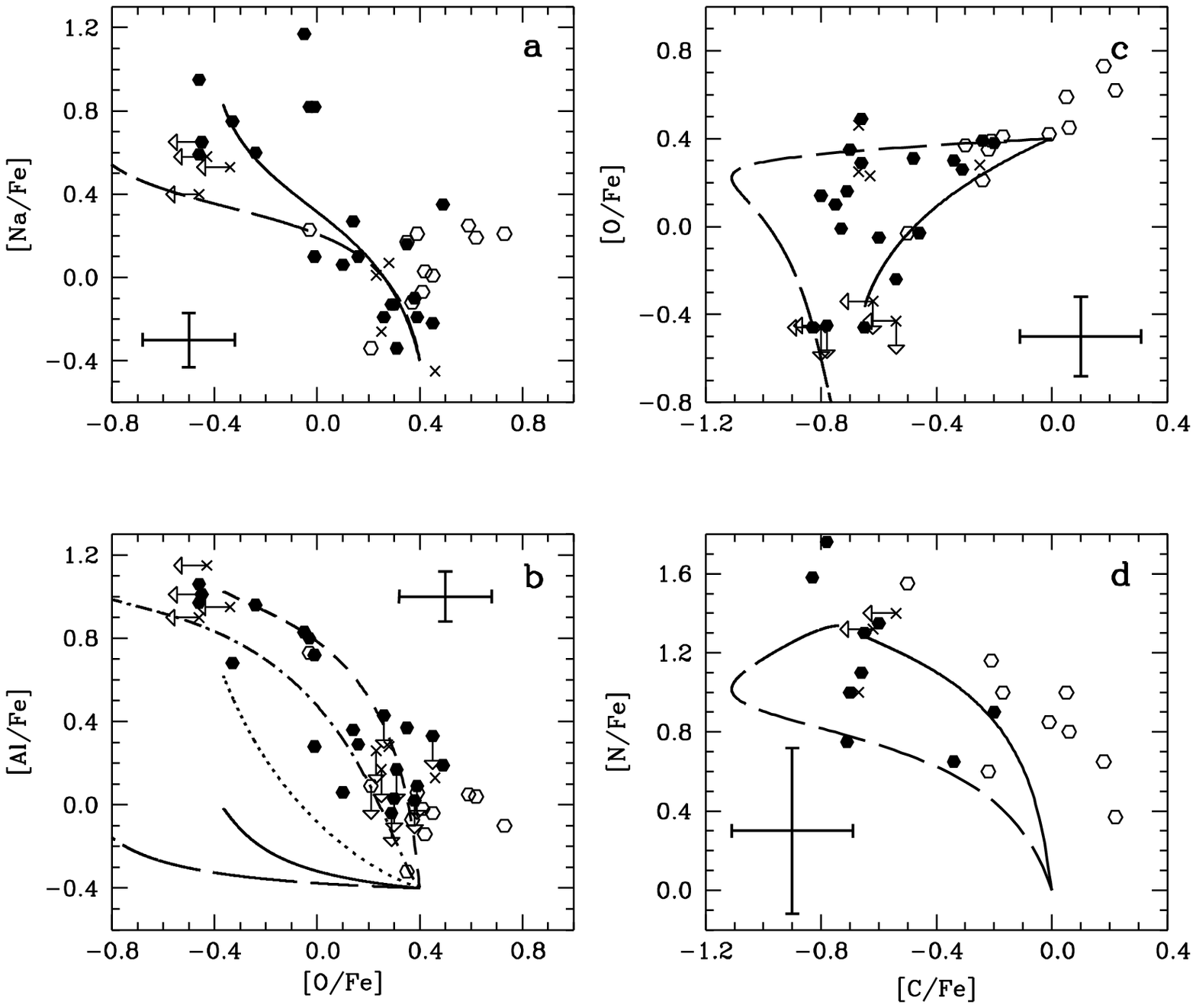}

\caption[]{The abundance trends seen in \omcen\ giants (symbols; from ND95)
are compared with the results of our deep mixing calculations for two sets of
mixing depth and rate (\delmix; \dmix, cm$^2$\,s$^{-1}$): (0.05; 5\,10$^8$)
-- solid, dotted and short-dashed lines, (0.06; 2.5\,10$^9$) -- long-dashed
and dot-short-dashed lines.  The former pertains to \omcen\ while the latter
corresponds to the best fit to the anticorrelation of [O/Fe] versus [Na/Fe]
in M\,13 (Fig.~3). In panel b the dotted line was calculated with an initial
abundance [$^{25}$Mg/Fe]\,=\,1.2, whereas the short-dashed and
dot-short-dashed lines were determined with [$^{25}$Mg/Fe] = 1.1 and the
$^{26}$Al$^{\mbox{\tiny g}}$(p,$\gamma)^{27}$Si reaction rate increased to
10$^3$ times the value given by CF88 (see discussion in Sect. 3.4). Open and
filled symbols refer to CO-strong and CO-weak stars, and crosses denote stars
with unidentified CO status, following ND95. In panel d the N abundances of
ND95 have been shifted by +0.5\,dex (for them to agree better with the data of
Brown \& Wallerstein (1993) and Norris \& Da Costa (1997))         }

\end{figure*}
%======================================================================

   Closer comparison of the results for \omcen\ and M~13 shows that the
apparent difference between the two clusters is driven by the more metal-rich
objects in \omcen, and we shall return to this point in Sect. 3.5.  It
suffices here to note that our conclusion on the need for deeper mixing in
the \omcen\ stars with very strong Na (which also have larger [Fe/H]) will
stand, because in more metal-rich objects the regions in which Na is
synthesized are closer to the HBS and less accessible for mixing than in the
more metal-poor ones.

   In ND95 (Sect 5.3.1) the presence of a ``floor'' to the carbon abundance
distribution at [C/Fe]$\sim -0.8$ was suspected.  This ``floor'' finds a
natural explanation in the present simulations: sufficiently deep mixing
which touches the C rise seen in Fig.~1 (below $\delta M$ = 0.15) (see comment
{\it (iii)} at the beginning of this section) cannot produce very large
surface carbon depletions (compare the solid and the long-dashed lines in
Fig.~2c).

   The deep mixing calculations for \omcen\ giants show that during the last
50\% of time spent by the model star between the onset of mixing and the core
helium flash the surface \cc\ ratio declines gradually from 9 to 6, which is
in good agreement with values 4--6 reported by Brown \& Wallerstein
(1989) for objects near the tip of the giant branch.

   The dot-long-dashed line in Fig.~3a presents results of our deep mixing
calculations performed with the new NeNa-cycle reaction rates (El Eid \&
Champagne 1995). Its form differs noticeably from that of the long-dashed
line calculated with the CF88 reaction rates, especially in the range
$0\leq$[O/Fe]$\leq 0.4$. This difference is entirely due to the considerably
higher new rate of the reaction \NeNa\ which causes a shift of the first rise
of the Na abundance far outwards from the O depleted layer (Fig.~4). As a
result, deep mixing first quickly produces a large Na enrichment at the
stellar surface and only then does the surface O abundance begin to
decline. Unfortunately, observational errors in the spectroscopic abundance
analysis do not allow us to make a definitive choice between the old and new
reaction rates from inspection of Fig.~3a.

   To summarize, the deep mixing scenario can explain evolutionary changes in
the abundances of C (and \cc\ ratios), N, O and Na.  It fails, however, to
interpret the anticorrelation of [O/Fe] vs. [Al/Fe] (Fig.~2b, solid line and
Fig.~3c, long-dashed line) and the correlation of [O/Fe] vs. [Mg/Fe]
(Fig.~3b, long-dashed line).

%==================== Fig.3 ===========================================
\begin{center}
\begin{figure}[h]
\hspace{3.0cm}\epsfxsize=10.1cm
\epsffile{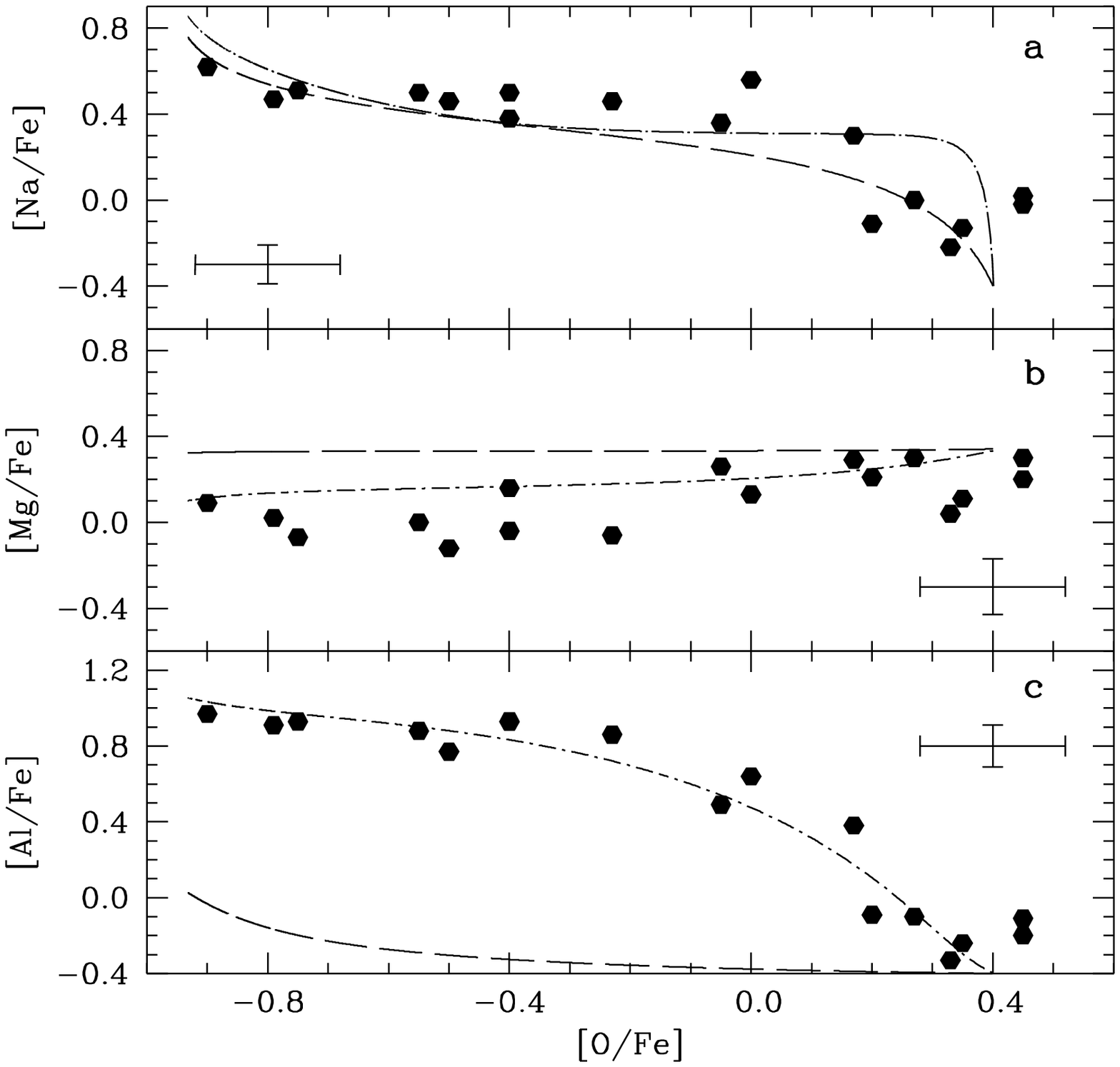}

\caption[]{The anticorrelations of [O/Fe] versus [Na/Fe] and [Al/Fe] and the
correlation of [O/Fe] versus [Mg/Fe] seen in M\,13 giants (symbols) compared
with the results of our deep mixing calculations for \delmix = 0.06 and \dmix
= 2.5\,10$^9$\,cm$^2$\,s$^{-1}$. Observational data are taken from Kraft et
al. (1997) with corrections of +0.05\,dex and --0.25\,dex applied by us to
their [Na/Fe] and [Al/Fe] values, respectively, to compensate for differences
between their adopted {\it gf} values and those of ND95. The dashed lines
were computed with standard input physics. The dot-long-dashed line in panel
a was calculated with the new NeNa-cycle reaction rates from El Eid \&
Champagne (1995), while the dot-short-dashed lines (panels b and c) were
calculated with the initial abundances [$^{24}$Mg/Fe] = 0 (as opposed to the
value +0.4 while we normally adopt), [$^{25}$Mg/Fe] = 1.1 and the
$^{26}$Al$^{\mbox{\tiny g}}$(p,$\gamma)^{27}$Si reaction rate increased to
10$^3$ times the CF88 value}

\end{figure}
\end{center}

%======================================================================

\subsection{When does the deep mixing start?}

   In the above considerations of both M\,13 and \omcen\ we started our deep
mixing calculations with a model star in which the HBS had just crossed the
H-He discontinuity left behind by the BCE at the end of the first
dredge-up. This and several neighbouring models are characterized by a small
drop in luminosity caused by the adjustment of the HBS to an increased fuel
supply when it encounters the H-He discontinuity.  Stellar evolution slows
down near this point and as a result a (subgiant) bump in the
globular-cluster luminosity function appears at the corresponding visual
magnitude. It was Sweigart \& Mengel (1979) who proposed that deep mixing
started only with that model because in preceding models there was a
molecular weight gradient between the HBS and the BCE built up during the MS
hydrogen burning which did not permit mixing (meridional circulation) to
operate. Observations of anomalous evolutionary changes of \cc\ and of the
$^3$He and $^7$Li abundances in evolved halo stars seem to support this idea
(Charbonnel 1995).  It is, however, clearly at odds with the observations in
clusters such as M92 which show depletions of carbon well below the level
postulated by Sweigart \& Mengel (Langer et al. 1986).  Moreover, recent
spectroscopic analysis of red giants in M\,13 (Kraft et al. 1997) has
revealed stars with enhanced Na and Al and depleted O (and even Mg!) at
luminosities low enough to draw the conclusion that at least in the giants of
M\,13 deep mixing (if it is responsible for the observed abundance
variations) begins earlier than Sweigart \& Mengel thought. Our comment on
this point is as follows.  Inspection of the chemical structure of our models
shows that the molecular weight gradient built up during the MS hydrogen
burning is {\it by far less steep} than the gradient at the point of deepest
penetration of the postulated deep mixing which has to be overcome in order
to dredge up freshly synthesized Al and even Na (if the reaction rates of
CF88 are used).  We therefore infer that in every globular cluster having red
giants with enhanced Al (and Na) and depleted O, even though we cannot
identify the process responsible, deep mixing began to operate well before
the point suggested by Sweigart \& Mengel.

A potential serious problem for deep mixing could be overcoming the H-He
discontinuity where, in canonical evolution, the mean molecular weight
experiences a jump within a very narrow mass interval as a result of the
deepest penetration of the convective envelope.  Since mixing signatures,
such as C-depletions, are seen at luminosities well below that at which
the HBS reaches this composition discontinuity (e.g.\ M92, Langer et al.\
1986), we must postulate that the mixing processes, perhaps acting in
concert with variable convective overshoot, prevent this 
discontinuity from arising (or smooth it out).  On the other hand, the
luminosity function bump that is predicted when the HBS burns through this 
discontinuity, is observed in many globular clusters (e.g.\ Alongi et al.\
1991).  Consequently, in these clusters, not all of the evolving red
giants can have had the discontinuity smoothed away.

%==================== Fig.4 ===========================================
\begin{center}
\begin{figure}
\hspace{3.0cm}\epsfxsize=10.1cm
\epsffile{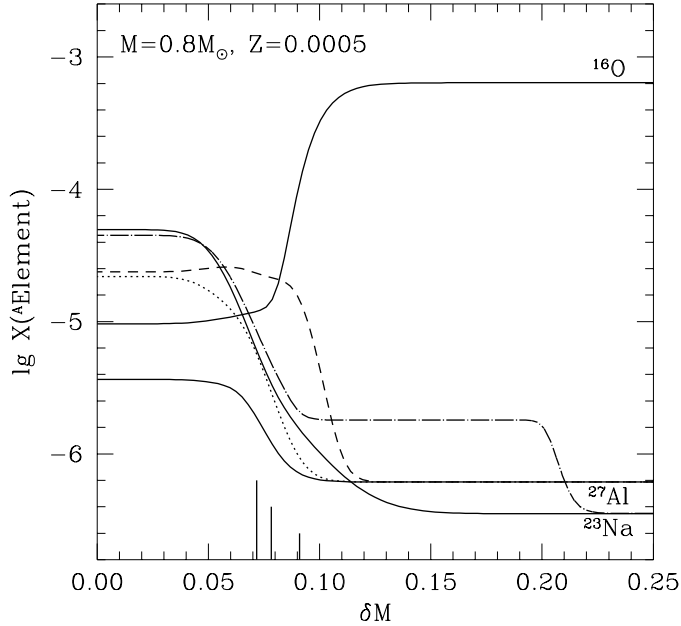}

\caption[]{Na and Al abundance distributions close to the HBS, calculated
under different assumptions: standard assumptions (see text) -- solid lines;
new NeNa-cycle reaction rates from El Eid \& Champagne (1995) --
dot-long-dashed line; initial abundance [$^{25}$Mg/Fe] = 1.2 -- dotted line;
[$^{25}$Mg/Fe] = 1.1 and the $^{26}$Al$^{\mbox{\tiny g}}$(p,$\gamma)^{27}$Si
reaction rate from CF88 multiplied by the factor 10$^3$ -- short-dashed line
}

\end{figure}
\end{center}

%======================================================================

\subsection{The Mg and Al abundances}

   As may be seen in Fig.~1, aluminium cannot be produced at the expense of
$^{24}$Mg in standard evolutionary calculations (see also comment {\it (ii)}
on Fig.~1 at the beginning of Sect.~3.2). On the other hand, $^{25}$Mg does
make Al (Fig.~1) but in amounts which are too small to explain the
observations in \omcen\ (see Fig.~2b, solid line) and M~13 (Fig.~3c,
long-dashed line).  Until recently one could speculate that in (some) GCRGs
the initial $^{25}$Mg abundance might be anomalously large as compared to its
scaled solar value (e.g. Langer \& Hoffman 1995; Paper~I). In the solar
chemical composition $^{24}$Mg is the most abundant magnesium isotope:
$^{24}$Mg/$^{25}$Mg/$^{26}$Mg = 79/10/11 (Anders \& Grevesse 1989). If one
assumes that [$^{25}$Mg/Fe]$>1.0$, then one can produce the observed Al
enhancements of about +1\,dex (see below).  Due, however, to an increased
contribution of $^{25}$Mg to the sum Mg = $^{24}$Mg+$^{25}$Mg+$^{26}$Mg and
the evolutionary transformation of this $^{25}$Mg into Al one would then
expect the total magnesium abundance [Mg/Fe] to decline with increasing
[Al/Fe].  Recently such an observational trend has been found in M\,13 giants
(Fig.~3b, symbols) by Shetrone (1996a) and Kraft et al.  (1997). On the other
hand, analysis of isotopic magnesium composition in a sample of 6 stars in
M\,13 (S96; Shetrone 1997) unexpectedly demonstrated that stars with
extremely large [Al/Fe] possessed noticeably reduced $^{24}$Mg and not
$^{25}$Mg ($\langle$[$^{24}$Mg/Fe]$\rangle = -0.33$ for 5 stars with the
largest [Al/Fe]).  It should be noted that the same analysis revealed (for
the first time in any star ever observed!) anomalous magnesium isotopic
ratios with an unusually increased contribution of the sum
$^{25}$Mg+$^{26}$Mg ([$^{25}$Mg+$^{26}$Mg/Fe] up to +0.21, average fractions
of the magnesium isotopes
$\langle^{24}$Mg$\rangle/\langle^{25}$Mg$\rangle/\langle^{26}$Mg$\rangle$ =
56/22/22).  Unfortunately, S96's analysis was not able to separate the
$^{25}$Mg and $^{26}$Mg isotopes and gave only their summed abundance, and
consequently, in deriving the magnesium isotopic ratios $^{25}$Mg and
$^{26}$Mg were assumed to have identical abundances.

   A straightforward interpretation of the anticorrelation of [Mg/Fe] versus
[Al/Fe] in the M\,13 giants, which guarantees an explanation of the O vs. Al
anticorrelations in the clusters M\,13 and \omcen\ and which also takes into
account the results of S96's isotopic analysis is to suppose that there is a
strong but still undetected low energy resonance in the reaction
$^{24}$Mg(p,$\gamma)^{25}$Al. The resonance has to provide this reaction with
a rate comparable with that of the $^{25}$Mg(p,$\gamma)^{26}$Al one in order
to ensure that the deep mixing reaches the depth where $^{24}$Mg is depleted
(see the $^{25}$Mg profile in Fig.~1). In this case aluminium would be a
product of the chain of reactions
$^{24}$Mg(p,$\gamma)^{25}$Al$(\beta^+\nu)^{25}$Mg(p,$\gamma
)^{26}$Al$^{\mbox{\tiny g}}$(p,$\gamma)^{27}$Si$(\beta^+\nu )^{27}$Al and the
beta-decay $^{26}$Al$^{\mbox{\tiny g}}(\beta^+\nu)^{26}$Mg together with the
channel $^{25}$Mg(p,$\gamma)^{26}$Al$^{\mbox{\tiny m}}(\beta^+\nu)^{26}$Mg
would take care of a large final abundance of the sum $^{25}$Mg+$^{26}$Mg
then dominated by $^{26}$Mg. ``Unfortunately'', nuclear physicists seem to
have little (if any) doubt concerning the current
$^{24}$Mg(p,$\gamma)^{25}$Al reaction rate (Arnould et al. 1995; Zaidins \&
Langer 1997).  Nevertheless, it would be interesting to know whether they can
{\it guarantee} that such a low energy resonance does not exist.  Of
course, any further experimental studies of the MgAl-cycle reaction rates
would be well worthwhile.

   An alternative is that in GCRGs with enhanced Na and Al and depleted O and
Mg abundances we actually observe products of hydrogen burning which has
occurred at much higher temperatures (say, at $T_6\equiv T/10^6\,\mbox{\rm
K}\sim 70$) than those reached in the HBS in the standard stellar models
($T_6\leq 55$).  We will explore this possibility in the next section.

   In the event that the above two suggestions cannot be realized, a third
and final possibility is to postulate that the extremely large Al
enhancements in GCRGs are a signature of an unusually overabundant initial
$^{25}$Mg. But now, to comply with the results of S96's magnesium isotopic
analysis, we have also to explain how stars with especially large initial
$^{25}$Mg abundances acquired a deficit in $^{24}$Mg.  In Sect.~4 we will
consider a primordial source which is able (in principle) to produce such an
abundance mixture and until then we merely assume that low mass stars in
M\,13 and \omcen\ had initially increased $^{25}$Mg.

   The dotted line in Fig.~2b (see also Fig.~4) shows how [Al/Fe] evolves
with [O/Fe] in \omcen\ giants in the case of an initial abundance
[$^{25}$Mg/Fe] = 1.2. Here we used the same values of \delmix\ and \dmix\ as
earlier because they had given good fits to the other three abundance trends
seen in \omcen.  Unfortunately, the new calculations disagree with
observations both quantitatively (there is not enough Al produced) and
qualitatively (the dotted line has a slope different from that hinted at by
the observations). The quantitative disagreement is caused by the fact that
with CF88 reaction rates the channel
$^{25}$Mg(p,$\gamma)^{26}$Al$(\beta^+\nu)^{26}$Mg dominates over the
$^{27}$Al producing channel (see above), and as a result, a large fraction of
$^{25}$Mg is wasted to synthesize $^{26}$Mg instead of $^{27}$Al. The wrong
slope of the theoretical dependence of [Al/Fe] on [O/Fe] comes from the
necessity of waiting until a large enough abundance of
$^{26}$Al$^{\mbox{\tiny g}}$ is built up for the chain of reactions
$^{26}$Al$^{\mbox{\tiny g}}$(p,$\gamma)^{27}$Si($\beta^+\nu)^{27}$Al to begin
competing with the beta-decay $^{26}$Al$^{\mbox{\tiny
g}}(\beta^+\nu)^{26}$Mg.  These undesirable effects can be diminished by
choosing a faster $^{26}$Al$^{\mbox{\tiny g}}$(p,$\gamma)^{27}$Si reaction
rate. Indeed, according to Arnould et al. (1995), for the range of
temperatures found in the HBS in GCRGs ($T_6\sim 40 - 50$) the
$^{26}$Al$^{\mbox{\tiny g}}$(p,$\gamma)^{27}$Si reaction rate may be
underestimated by a large factor of $\sim 10^3$ in CF88.  If this
acceleration factor is applied, the $^{27}$Al producing channel gets so wide
that now one can even use a somewhat smaller value of the initial $^{25}$Mg
abundance. In Fig.~2b (see also Fig.~4) the short-dashed line was calculated
with [$^{25}$Mg/Fe] = 1.1 and the $^{26}$Al$^{\mbox{\tiny
g}}$(p,$\gamma)^{27}$Si reaction rate equal to $10^3$ times its CF88 value.
The dot-short-dashed lines in Figs.~2b and 3c were calculated under the same
assumptions but for the mixing depth and rate required by the O vs. Na
anticorrelation in M\,13. Now theory matches the observations! The initial
$^{24}$Mg abundance has no influence on the results of these calculations. In
Fig.~3b the dot-short-dashed line is obtained assuming the initial abundance
[$^{24}$Mg/Fe] = 0 (instead of +0.4) which, as expected, produces a decline
of the total magnesium abundance [Mg/Fe] with increasing [Al/Fe] due to
consumption of the now abundant $^{25}$Mg. It seems, however, that
observations demand even a larger (initial) deficit in $^{24}$Mg.

   The adoption of [$^{24}$Mg/Fe]\,=\,0, [$^{25}$Mg/Fe]\,=\,1.1 and
[$^{26}$Mg/Fe]\,=\,0 results in [$^{25}$Mg+$^{26}$Mg/Fe]\,=\,0.81, [Mg/
Fe]\,=\,0.33 and $^{24}$Mg/$^{25}$Mg/$^{26}$Mg\,=\,37/58/5, i.e. a $^{25}$Mg
dominated mixture of the magnesium isotopes; and for the deep mixing
presumably operating in M\,13 giants (\delmix\,=\,0.06,
\dmix\,=\,2.5\,$10^9$\,cm$^2$\,s$^{-1}$) just before the core helium flash we
find [$^{24}$Mg/Fe]\,=\,0.004, [$^{25}$Mg/Fe]\,=\,\mbox{--0.86},
[$^{26}$Mg/Fe]\,=\,0.61 and, consequently, [$^{25}$Mg+$^{26}$Mg/Fe] = 0.35,
[Mg/Fe]\,=\,0.10 and $^{24}$Mg/$^{25}$Mg/$^{26}$Mg\,=\,64/1/35. Since,
however, the observations {\it cannot separate} $^{25}$Mg and $^{26}$Mg this
final mixture is not distinguished from the one $^{24}$Mg/$^{25}$Mg/$^{26}$Mg
= 64/18/18 which has the ratios close to the average ones reported by S96.

   Shetrone interpreted his results of the magnesium isotopic analysis in
terms of the $^{25}$Mg and $^{26}$Mg abundances remaining unchanged during
deep mixing. But as we have seen there are two alternative interpretations:
{\it (i)} initially the magnesium isotope mixture consisted of almost pure
$^{24}$Mg (as, for instance, in the M\,13 giant L598 of S96's sample which
has $^{24}$Mg/$^{25}$Mg/$^{26}$Mg = 94/3/3, and which shows little evidence of
mixing); if, in addition, there is a strong low energy resonance in the
reaction $^{24}$Mg(p,$\gamma)^{25}$Al, then after mixing we will have
$^{24}$Mg depleted and both the sum $^{25}$Mg+$^{26}$Mg and $^{27}$Al,
enhanced; {\it (ii)} initially $^{24}$Mg was depleted and $^{25}$Mg
substantially increased (primordially!); during deep mixing the $^{24}$Mg
abundance remains constant and that of $^{27}$Al increases at the expense of
$^{25}$Mg, the sum $^{25}$Mg+$^{26}$Mg being dominated by $^{26}$Mg which is
also produced from $^{25}$Mg. In both cases one obtains approximately
what Shetrone has observed.

%==================== Fig.5 ===========================================

\begin{figure*}
\epsfxsize=16cm
%\epsffile [80 140 510 520] {f5.eps}
\epsffile [80 120 510 520] {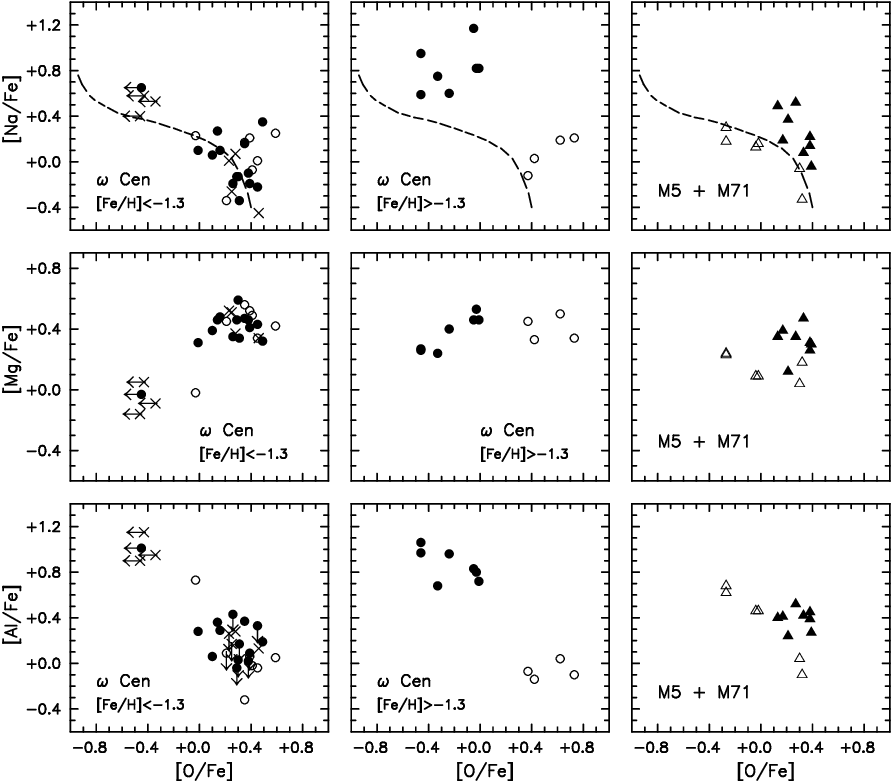}

\caption[]{The dependence of [Na/Fe], [Mg/Fe] and [Al/Fe] on [Fe/H] for
(left) \omcen\ giants with [Fe/H] $<$ --1.3, (middle) \omcen\ stars with
[Fe/H] $>$ --1.3, and (right) M5 and M71 giants.  For \omcen\ the symbols are
as defined in Fig.~2, while for M5 (open triangles) and M71 (filled
triangles) the data have been taken from Shetrone (1996a) corrected by +0.05
and --0.25 for Na and Al as in Fig.~3.  The dashed lines are the deep
mixing simulation for Na in  M~13 shown in Fig.~3a}

\end{figure*}
%======================================================================

\subsection{Metallicity dependent effects in \omcen }

   We noted in Sect. 3.2 that the more metal-rich stars in \omcen\ were in
large part responsible for the apparent differences between it and M~13.  In
Fig.~5 we divide the \omcen\ sample into two abundance groups: on the left
are objects having [Fe/H] $<$ --1.3, and in the middle those with [Fe/H] $>$
--1.3.  On the right, for comparison purposes, we also show data for the
metal-richer globular clusters M5 ([Fe/H] = --1.2) and M71 ([Fe/H] = --0.8)
from Shetrone (1996a).  To focus further discussion we also superimpose on the
diagrams the model fits for Na in M~13 (the dashed line from Fig.~3a).

   Inspection of the figure reveals several interesting points.  First,
concerning Na, one sees that the metal-poorer objects in \omcen\ have an
[Na/Fe] versus [O/Fe] correlation similar to that seen in M~13.  In contrast,
however, the metal-rich \omcen\ stars have [Na/Fe] values which are larger by
$\sim$ 0.4 dex than in M~13.  Such large Na enhancements do not exist in the M5
and M71 sample, which has comparable [Fe/H].

   Second, concerning Mg, we see among the metal-poor \omcen\ group the same
[Mg/Fe] versus [O/Fe] anticorrelation as discussed by Shetrone (1996a) and
Kraft et al. (1997) for M13.  The spread in [Mg/Fe] ($\sim$ 0.5--0.6), appears
to be slightly larger that observed for M~13 in Fig.~3 ($\sim$ 0.3),
suggestive of more extreme mixing in \omcen.  This is, however, very
different from the behaviour of [Mg/Fe] versus [O/Fe] in the metal-rich
\omcen\ stars, where one sees little evidence for a spread in [Mg/Fe].

    Finally, both groups of \omcen\ stars appear to exhibit the same [Al/Fe]
versus [O/Fe] behavior.  A concomitant, important, point is that there is
little if any dependence of [Mg/Fe] on [Al/Fe] in the metal-rich \omcen\
giants, in stark contrast to what is found in the metal-poor \omcen\ and M~13
giants!

   One may summarize the above results by stating that the mixed metal-rich
giants in \omcen, in comparison with their metal-poorer counterparts, have
similar Al enhancements, larger Na enhancements, and no accompanying Mg
depletion.  The metal-rich \omcen\ stars clearly offer an important
constraint on our understanding of the manner in which they have been
enriched.  For those who would take up the challenge we provide the following
information and caveats.  The mixed objects in question are ROA 150, 162,
231, 248, 357, 371 and 480, and were included in the work of ND95 in an
extremely biased manner as examples of objects having the most extreme
abundance peculiarites in the cluster.  All have enhancements of the
s-process elements, which ND95 attribute to primordial enhancement from low
mass AGB stars.  They have no counterpart in the more metal-rich clusters,
such as M5 and M71.
   
%======================================================================

%==================== The primordial scenario =========================
\section{The primordial scenario}

   In the primordial scenario, star-to-star abundance variations in globular
clusters are thought to be produced before or during formation of low mass
stars. Possible sources of primordial pollution include winds from massive
stars, SNe explosions, winds and planetary nebulae (PNe) ejection by AGB
stars and novae. Are any traces of nucleosynthesis yields from the earlier
generations of stars seen in globular clusters?  If one assumes that
protoglobular clusters formed from material having the big bang chemical
composition with zero heavy element content, then the answer is definitely
yes, since in present-day clusters stars show a wide spectrum of elements
heavier than helium, and one is forced to seek such primordial sources.

   The first objects to contribute to the self-enrichment of globular
clusters are massive stars during hydrostatic mass loss and on their final
evolutionary stage -- SNeII (we do not consider the hypothesized Very Massive
Objects). The modern radiation-driven wind theory (Kudritzki et al. 1991;
Jura 1986) predicts, however, that during hydrostatic evolution massive stars
with Z = 0 will not lose a considerable amount of their mass (Maeder
1992). Therefore, the only nucleosynthesis yields from the first generation
of massive stars are most probably those produced by SNeII and consist mainly
of O (and other $\alpha$-elements) (Woosley \& Weaver 1995). It is also
important to note that as argued by Laird \& Sneden (1996) and Carney
(1996) there is no evidence for any {\it progressive enrichment} of
clusters by the SNeIa responsible for the production of the iron-peak
elements in Galactic chemical evolution (Timmes et al. 1995, hereafter TWW95)
and which chronologically follow SNeII.

   The next potential polluters of the intracluster medium are AGB
stars. ND95 reported that abundances of the heavy neutron-addition elements,
which are presumably produced in the s-process nucleosynthesis in $1 -
3\,M_\odot$ AGB stars rise as [Fe/H] increases in \omcen\ red giants.  Since
intermediate mass stars ($3 - 8\,M_\odot$) evolve faster than $1 -
3\,M_\odot$ objects, one expects that some products of nucleosynthesis from
intermediate mass AGB stars could also be present in the material captured by
contracting low mass stars in globular clusters or accreted by them later.

\subsection{Nucleosynthesis yields from SNe}

   The chemical enrichment history of globular clusters seems to be quite
different from the chemical evolution of the bulk of the Galaxy. One of the
plausible scenarios of globular cluster formation was proposed by Cayrel
(1986) and elaborated upon further by Brown, Burkert \& Truran (1991,
1995). In this scenario massive stars form first in a protocluster's dense
core from material having the big bang composition. They evolve rather
quickly (on a time scale of $\sim 10^6 - 10^7$ years) and explode as SNeII
without leaving any remnants. A supershell produced by shocks from the
multiple SNe explosions sweeps up and compresses the protocluster material
which now consists of a mixture of the big bang and SNeII ejecta
compositions. It is this supershell that becomes the birthplace for stars
covering the whole mass spectrum. There are a number of arguments, both
theoretical and observational, supporting this scenario (see the papers cited
above). One of these is the absence of low mass stars having the primordial
big bang composition. We shall not, however, discuss here all advantages and
disadvantages of different globular clusters formation models. We merely
emphasize that both intermediate and low mass stars in globulars were very
likely to form out of material polluted by SNeII. Two things are necessary to
specify the element abundances in this material: first, mass dependent
nucleosynthesis yields from the SNeII whose progenitors had the big bang
initial composition, and second, an estimate of dilution of these yields in
the supershell.

   In Fig.~6 the abundances of some of the SNeII ejecta which are important
for our work are presented as functions of the initial mass of the SN
progenitor (which is also its final mass because in these calculations
stellar winds were not taken into account) for $Z=0$ for the models of
Woosley \& Weaver (1995) as adopted by TWW95 (their Sect. 2.4).  The large
variations of abundances from star to star are mainly caused by the
uncertainty in modelling the explosion and the sensitivity of models to the
interaction between various convective zones during the late stages of the
evolution. Despite these large random-like abundance variations TWW95 argue
that after convolution with an appropriate initial mass function and
integrating over time one gets quite reasonable results, and indeed they do
succeed in reproducing the observed evolutionary changes of abundances for a
large number of elements lighter than zinc in a simple model of galactic
chemical evolution.

%==================== Fig.6 ===========================================
\begin{center}
\begin{figure}
\epsfxsize=10.1cm
\hspace{3.0cm}\epsffile{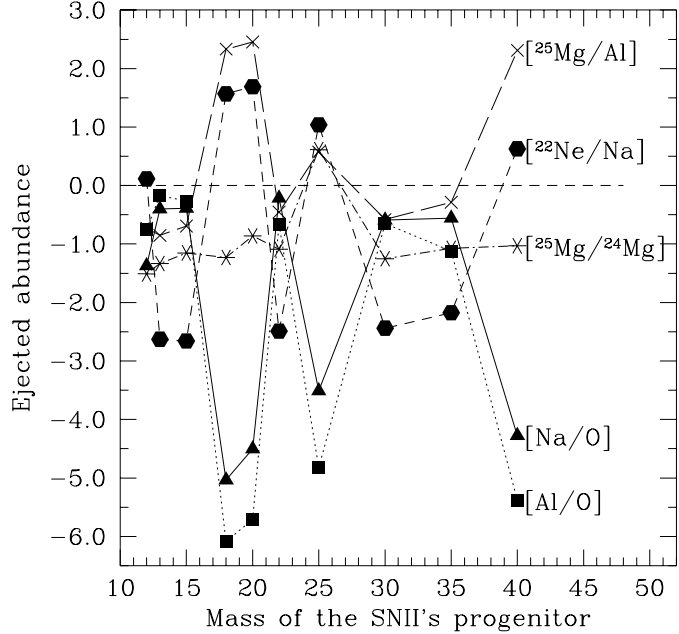}

   \caption[]{Abundances of some nuclides ejected by SNeII (following Woosley
\& Weaver 1995)}

\end{figure}
\end{center}

%======================================================================

   The range of masses of the SN progenitors in Fig.~6 is $12\leq
M/M_\odot\leq 40$. Following TWW95 and Cayrel (1986) we assume that the
initial mass function in the protoglobular cluster's dense core was a low
mass cutoff Salpeter (1955) power law $\Psi(M)\propto M^{-(1+x)}$ with a
slightly different exponent of $x=-1.31$ (instead of --1.35) which gave the
best fit to the observed element evolution. After convolution of the
abundances from Fig.~6 with $\Psi(M)$ over the range 12 -- 40 M$_\odot$ we find
the average abundances ejected by SNeII.  The latter are diluted in the
supershell by material having the big bang composition. The dilution
coefficient can be estimated as the ratio $v_s/v_{ej}\sim 10^{-3}$ of the
local intracluster sound velocity $v_s\sim 10$\,km\,s$^{-1}$ to the initial
velocity of the SN's ejecta $v_{ej}\sim 10^4$\,km\,s$^{-1}$ (Cayrel
1986). This estimate neglects the adiabatic phase of the SN's expansion and
takes into account only the ``snowplow'' phase. In Table~1 the final
abundances expected as a result of SNeII explosions and subsequent dilution
in the supershell are presented.

%======================================================================
\begin{table}
\caption[]{Abundances expected as the result of SNeII explosions
          }
\begin{center}
\begin{tabular}{ll}
\noalign{\smallskip}
\hline
\noalign{\smallskip}
abundance(s) & SNeII \\
\noalign{\smallskip}
\hline
\noalign{\smallskip}
\mbox{[C/Fe]} & --0.25 \\
\mbox{[N/Fe]} & --2.44 \\
\mbox{[O/Fe]} & --0.05 \\
\mbox{[$^{20}$Ne/Fe]} & +0.10 \\
\mbox{[Na/Fe]} & --0.52 \\
\mbox{[$^{24}$Mg/Fe]} & --0.05 \\
\mbox{[$^{25}$Mg/Fe]} & --1.25 \\
\mbox{[$^{26}$Mg/Fe]} & --1.27 \\
\mbox{[Mg/Fe]} & --0.15 \\
\mbox{[Al/Fe]} & --0.67 \\
\mbox{[Si/Fe]} & --0.15 \\
\mbox{[Fe/H]} & --2.31 \\
\mbox{[$^{22}$Ne/Na]} & --2.34 \\
\mbox{[$^{25}$Mg+$^{26}$Mg/Al]} & --0.59 \\
\mbox{$^{24}$Mg/$^{25}$Mg/$^{26}$Mg} & 98/1/1 \\
\noalign{\smallskip}
\hline
\noalign{\smallskip}
\end{tabular}
\end{center}

\end{table}
%======================================================================

   From these data we draw the following conclusions: {\it (i)} the value of
[Fe/H] corresponds to that of the most metal-deficient globular clusters;
{\it (ii)} the values of [$\alpha$/Fe] are lower than observed in globular
clusters which results, in large part, from the overproducton of Fe in the
model supernovae (see TWW95); {\it (iii)} N is extremely underabundant; {\it
(iv)} the $^{22}$Ne/Na ratio is far less than the initial one assumed in
GCRGs (Sect.~3.1); {\it (v)} abundances of $^{25}$Mg and $^{26}$Mg are very
low compared to $^{24}$Mg and Al. Inspection of Fig.~6 shows that the
conclusion concerning the very low $^{25}$Mg/$^{24}$Mg ratio in the
primordial mixture is rather robust.  We thus have to look for other
nucleosynthesis source(s) of $^{22}$Ne, $^{25}$Mg and $^{26}$Mg. TWW95
proposed the intermediate mass AGB stars as the main producers of the
$^{25}$Mg and $^{26}$Mg isotopes. Our calculations support this idea (see
below).  As regards N, TWW95 reported that its abundance in SNeII ejecta can
be much larger if one takes into account some convective overshoot in the low
metallicity SNII's progenitor but they also infer that AGB stars can
contribute much to the galactic N (and C) production.

\subsection{Yields from intermediate mass AGB stars}

   To take into account the contribution to galactic chemical evolution from
intermediate mass AGB stars TWW95 made use of the results of parameterized
nucleosynthesis calculations by Renzini \& Voli (1981) who, however, followed
only the evolution of abundances of $^{12}$C, $^{13}$C, $^{14}$N and $^{16}$O
for $Z\geq 0.004$.  Our calculations supplement those of Renzini \& Voli in
two respects.  First, we consider also the evolution of elements heavier than
O and, second, we choose $Z$ as low as 10$^{-4}$ and allow different relative
distributions of nuclides within $Z$ which turns out to have a pronounced
effect on the final abundances.

   As a representative of intermediate mass AGB stars we considered a
$5\,M_\odot$ object. Thermal pulses of the helium burning shell started with
the core mass $M_{\mbox{\tiny c}}=0.96\,M_\odot$. It should be emphasized
again that we did not follow the AGB stellar evolution.  Instead we
considered the parameterized nucleosynthesis in the $5\,M_\odot$ AGB star
following the description of Renzini \& Voli (1981) (for more details see
Paper~II). The core mass is constrained to remain smaller than the
Chandrasekhar limit $M_{\mbox{\tiny Ch}}\sim 1.4\,M_\odot$, but long before
$M_{\mbox{\tiny c}}$ could approach $M_{\mbox{\tiny Ch}}$ some kind of
instability is believed to force envelope ejection in the form of a PN which
terminates AGB evolution (Renzini 1981; Wagenhuber \& Weiss 1994). The exact
upper limit for the number of pulses before the PN's ejection is not
known. It can be approximately constrained by the observed relation between
white dwarf masses and the initial masses of their MS progenitors. Such
relations show that intermediate mass stars with lower metallicity may
survive longer on the AGB (Weidemann 1987). We have chosen $N=400$ as the
limiting number of pulses in our low metallicity AGB nucleosynthesis
calculations. This number results in the final core mass $M_{\mbox{\tiny
c}}=1.12\,M_\odot$. We are aware that the chosen value of $N$ may be
overestimated, but smaller values ($N=100-200$) give the same qualitative
(but of course less pronounced quantitative) results.

%==================== Fig.7 ===========================================
\begin{center}
\begin{figure}
\epsfxsize=10.1cm
\hspace{3.0cm}\epsffile{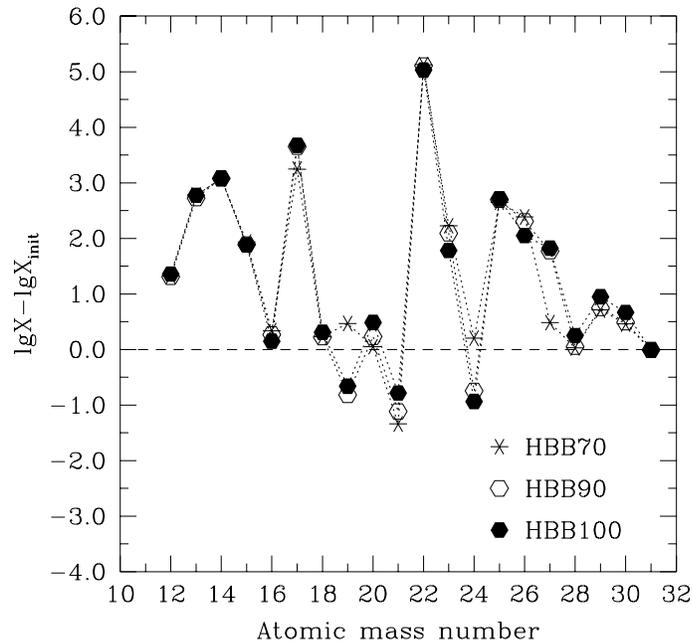}

\caption[]{Nucleosynthesis yields of some light nuclides from intermediate
mass AGB stars after 400 pulses. The notation HBBT6 signifies that HBB was
assumed to occur at the temperature T6\,10$^6$\,K.  The atomic mass number 26
corresponds to $^{26}$Mg.  The initial chemical composition was that given in
Table~1
          }
\end{figure}
\end{center}

%======================================================================

   In Fig.~7 the final (i.e. after 400 pulses) surface abundances in the
envelope of the $5\,M_\odot$ AGB star are plotted for three different hot
bottom burning (HBB) temperatures: $T_{\mbox{\tiny HBB}}=70, 90$ and
100\,10$^6$\,K.  It should be noted that the model value of $T_{\mbox{\tiny
HBB}}$ is also very uncertain because it strongly depends on the depth within
the HBS which can be reached by the BCE during the interpulse period, with
the depth being very sensitive to the poorly known extent of convective
overshoot (Lattanzio \& Frost 1997). Unfortunately, it is $T_{\mbox{\tiny
HBB}}$ that determines whether $^{24}$Mg is transformed into Al at the BCE
(Fig.~7). The initial chemical composition for these calculations was that
given in Table~1 (see Sect.~4.1).

   From Fig.~7 we infer that if HBB is neglected (this approximately
corresponds to the results shown by asterisks for $T_{\mbox{\tiny
HBB}}=70\,10^6$\,K) the main new results (as compared to those obtained by
Renzini \& Voli (1981)) are considerable increases of the $^{22}$Ne/Na ratio
and of the $^{25}$Mg and $^{26}$Mg abundances. The $^{22}$Ne, $^{25}$Mg and
$^{26}$Mg isotopes are primarily produced by $\alpha$-capture reactions
whereas Na is synthesized in the reactions
$^{22}$Ne(n,$\gamma)^{23}$Ne($\beta^-\overline{\nu})^{23}$Na.  Aluminium is
not produced during pulses because there are no $\alpha$-capture reactions
which result in Al, on the one hand, and the neutron capture cross section
for the reaction $^{26}$Mg(n,$\gamma)^{27}$Mg followed by the beta-decay
$^{27}$Mg($\beta^-\overline{\nu})^{27}$Al is very low ($\sigma_{\mbox{\tiny
n}\gamma}=0.084$\,mb), on the other.

   In Table~2 we compare the abundances of $^{22}$Ne, Na, $^{25}$Mg and
$^{26}$Mg which are achieved in the H-He intershell just after the 400th
pulse for three initial compositions: (1) solar; (2) $Z=10^{-4}$ and the
relative abundance distribution as described in Sect.~3.1; (3) abundances
from Table~1. We see that especially large increases of the ratio
$^{22}$Ne/Na and of the $^{25}$Mg and $^{26}$Mg abundances are obtained for
the third mixture, which was derived from the globular cluster
self-enrichment model.

%======================================================================
\begin{table}
\caption[]{Abundances ($\lg X/X_{\mbox{\tiny i}}$) in the H-He intershell 
           of the $5\,M_\odot$ AGB star after the 400th pulse
          }
\begin{center}
\begin{tabular}{lllll}
\noalign{\smallskip}
\hline
\noalign{\smallskip}
chem. comp. & $^{22}$Ne & Na & $^{25}$Mg & $^{26}$Mg \\
\noalign{\smallskip}
\hline
\noalign{\smallskip}
solar & 2.06 & 1.57 & 1.40 & 1.05 \\
$Z=0.0001^a$ & 2.36 & 2.19 & 1.69 & 1.20 \\
SNeII$^b$ & 4.80 & 1.79 & 2.51 & 1.98 \\
\noalign{\smallskip}
\hline
\noalign{\smallskip}
\end{tabular}
\end{center}

(a) the relative distribution of heavy nuclides within $Z$ is described
in Sect.~3.1; (b) the initial composition from Table~1.

\end{table}
%======================================================================

   In Paper~II our estimate of the fraction of material which first takes
part in the nucleosynthesis processes in intermediate mass AGB stars and is
later captured by low mass stars $q$ (the dilution coefficient) was $0.1 -
0.2$. This estimate assumes homogeneous distribution of the processed
material among low mass stars. While this assumption appears to be good one
for the nucleosynthesis yields from SNe because the post-shock turbulence is
thought to mix the protocluster material well (Brown et al.  1995), yields
from AGB stars are most likely to be distributed inhomogeneously and
primarily captured by low mass stars nearest to them during ejection of their
envelopes. In this case the coefficient $q$ may be even larger.
   
   The discussion given above supports the idea that intermediate mass AGB
stars could be a source of the increased initial $^{25}$Mg (and $^{26}$Mg)
abundance in GCRGs. Indeed, it follows from Fig.~7 that a value of $q\geq
0.2$ would be quite enough to increase the $^{25}$Mg abundance from
[$^{25}$Mg/Fe] = \mbox{--0.59} (Table~1) up to [$^{25}$Mg/Fe]$\geq
1.0$ as required by our deep mixing calculations (Sect.~3.4).  At the same
time this would bring the $^{22}$Ne/Na ratio close to the value
[$^{22}$Ne/Na] = 0 which makes possible the synthesis of Na in GCRGs inferred
to occur at the expense of $^{22}$Ne.

   Our calculations of the nucleosynthesis yields from the $5\,M_\odot$ AGB
star confirm the conclusion of TWW95 that AGB stars can contribute to the
enrichment of the interstellar medium (and in our case of low mass stars in
globular clusters) in C and N. We also repeat our result from Paper~II that
the intermediate mass AGB stars can be responsible for some primordial
enrichment of low mass stars in Na and Al (Fig.~7).  Contrary, however, to
the situation for Na, which is produced during pulses, Al is produced from
$^{24}$Mg only during HBB, the details of which are not fully constrained by
theory.  At temperatures higher than about $70\,10^6$\,K the
$^{24}$Mg(p,$\gamma)^{25}$Al reaction goes faster than proton captures by
$^{25}$Mg and $^{26}$Mg (CF88), the latter two isotopes being also produced
copiously during pulses. As a result of very hot bottom burning
($T_{\mbox{\tiny HBB}}\geq 90\,10^6$\,K) which is predicted by recent
evolutionary calculations (Lattanzio et al. 1997), the $^{25}$Mg enhancement
can be accompanied by a deficit of $^{24}$Mg and by an increase of Al. In
order, however, to produce low mass stars with even a twofold decrease of the
initial $^{24}$Mg abundance we need a dilution coefficient as large as
$q=0.5$ and of course in addition we get some primordial Al enhancement. It
should be noted that the abundance of O is not reduced as the result of HBB
in intermediate mass AGB stars (Fig.~7) since O is synthesized from C during
pulses. Deep mixing in red giants is therefore still required.

   To summarize, the advantages of the proposed primordial scenario are the
following: {\it (i)} it supplies low mass stars with a large initial
abundance of $^{25}$Mg; {\it (ii)} it explains why in GCRGs with especially
large Al enhancements some $^{24}$Mg depletion is also observed (because low
mass stars with a low initial $^{24}$Mg abundance are expected to possess a
large initial $^{25}$Mg); {\it (iii)} it accounts for some C (and N)
primordial enrichment of low mass stars as is observed in globular
clusters (Fig.~2, open symbols). Its apparent deficiencies are: {\it (i)} it
assumes rather large dilution coefficients ($q\geq 0.5$) to comply with the
observed low abundance [$^{24}$Mg/Fe] in M\,13 giants; {\it (ii)} it seems to
disagree with the approximate constancy of the sum C+N+O reported in several
globular clusters (and noted in Sect. 1) because it assumes the initial
abundance of C (and N) in low mass stars to be a function of the dilution
coefficient $q$, which may change from star to star; {\it (iii)} and,
unfortunately, there are still many uncertainties in this scenario which do
not allow us to draw more definite conclusions.

   We conclude this section by noting that our calculations of the s-process
nucleosynthesis in the $5\,M_\odot$ AGB star with $Z=10^{-4}$ (which are in
excellent agreement with the earlier results of Busso et al. (1988)) show
that there is no substantial production of neutron-addition nuclides.  Hence,
we do not expect that the increased initial $^{25}$Mg abundance in GCRGs has
to correlate with any enhancement of the s-process elements. At present
s-process nucleosynthesis is believed to occur in low mass ($1 - 3\,M_\odot$)
AGB stars where the neutron source reaction is $^{13}$C($\alpha$,n)$^{16}$O
rather than $^{22}$Ne($\alpha$,n)$^{25}$Mg as in our case (Gallino et
al. 1988).  Therefore, the observational data of ND95 showing that in \omcen\
giants abundances of some typical s-process elements (Y, Ba, La and Nd) rise
as [Fe/H] increases was interpreted by them as evidence of primordial
enrichment by $1 - 3\,M_\odot$ AGB stars.  Or to reverse the argument, if one
accepts that the s-process elements are enhanced in \omcen\ by the ejecta
of $1 - 3\,M_\odot$ AGB stars, it follows that this should be accompanied
by overabundances of the heavy nuclides of Mg resulting from the ejecta of
their $3 - 8\,M_\odot$ counterparts.

\subsection{A ``black box'' solution}

   If the observed Al enhancements in red giants with depleted [Mg/Fe] in the
clusters M\,13 and \omcen\ are in fact produced at the expense of $^{24}$Mg
and not $^{25}$Mg and if nuclear physicists confirm the currently-accepted
rate of the reaction $^{24}$Mg(p,$\gamma)^{25}$Al, the only remaining
explanation of the MgAl anticorrelation is hydrogen burning at much higher
temperatures than those ($T_6\leq 55$) found in the HBS in the standard
evolutionary models.

   To test this idea we have considered hydrogen burning at constant
temperature and density. The density was chosen as $\rho =44.7$\,g\,cm$^{-3}$
as in the deep mixing study of Langer et al. (1993) and the initial chemical
composition was that specified in Sect.~3.1. Nucleosynthesis calculations
were interrupted when 5\% of hydrogen was consumed.  After that the
calculated abundances were mixed with unprocessed material whose fraction was
varied from 0\% to 100\%. Temperature was treated as a free parameter. The
results are shown in Fig.~8 where the lines are the computed correlations
between the final abundances of O, Na, Mg and Al. Crosses on the solid lines
correspond (from right to left) to mixtures in which the fraction of
unprocessed material is $q=1.0, 0.9, 0.8,$ \ldots (the last crosses seen on
the left on the solid lines have $q=0.1$). The range of temperatures fitting
the anticorrelations of [O/Fe] versus [Na/Fe] and [Al/Fe] and the correlation of
[O/Fe] versus [Mg/Fe] has turned out to be strikingly narrow: $T_6=70$
(short-dashed line), $T_6=74$ (solid line) and $T_6=78$ (dot-short-dashed
line). Hence, were this idea right we could very precisely estimate the
temperature of the hydrogen burning whose products are seen in GCRGs:
$T_6=74\pm 2$!  The choice of density does not affect this estimate and if we
take the amount of H to be consumed considerably different from 5\% then it
becomes more difficult to fit all three observed abundance correlations with
the same value of temperature. It is interesting that a very similar result
(with a slightly higher temperature $T_6=78$) is obtained when one considers
hydrogen burning in a massive convective core (we have modelled the core
structure with a polytrope $n=1.5$ and this time the free parameter to adjust
was the central temperature).

%==================== Fig.8 ===========================================
\begin{center}
\begin{figure}
\epsfxsize=10.1cm
\hspace{3.0cm}\epsffile{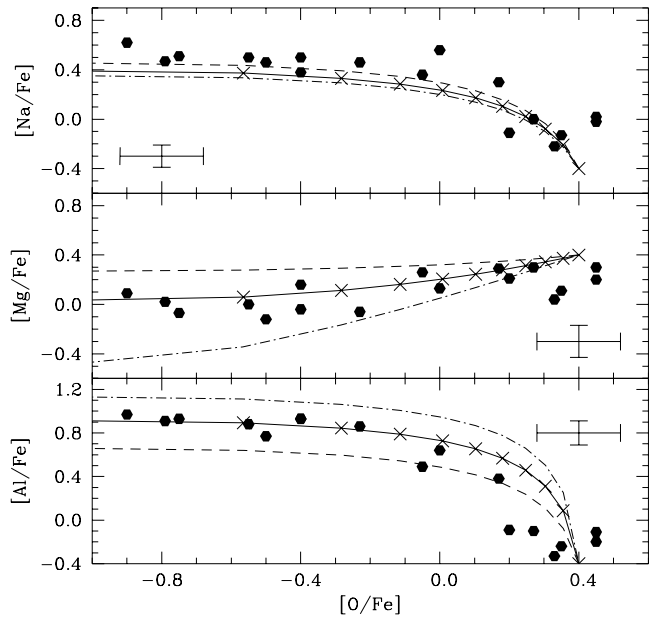}
   
\caption[]{Relations between the abundances of O, Na, Mg and Al (lines) in
mixtures with the fraction of unprocessed material varied from 0\% to 100\%
(crosses correspond to 100\%, 90\%, 80\%,\ldots\ from right to left). In the
processed material hydrogen burning at constant density
$\rho=44.7$\,g\,cm$^{-3}$ and temperature has been followed until 5\% of H
was consumed. The temperature has been adjusted ($T_6=70$ -- short-dashed
lines, $T_6=74$ -- solid lines and $T_6=78$ -- dot-short-dashed lines) to fit
the correlations in M\,13 giants (symbols). The initial chemical composition
was that described in Sect.~3.1}

\end{figure}
\end{center}
%======================================================================

   The next question to answer is which stellar environment may be identified
with the ``black box'' described above. We calculated a ZAMS model of a
$M=125\,M_\odot$ star with $Z=4\,10^{-4}$ but found that it had a central
temperature $T_6=53$, which is too low. Any primordial origin of the
hypothesized ``black box'' meets the difficulty of explaining why after
consumption of only 5\% of H the material was ejected into the intracluster
medium.  Another problem is understanding how some low mass stars succeeded
in capturing as much as 90\% of the material ejected by the ``black box''
(Fig.~8). These two problems are, however, easily solved if we place such a
``black box'' inside a star ascending the RGB, bearing in mind, of
course, that we now have to think of a mechanism which can increase the
temperature in the HBS up to the value $T_6=74$. Recently Langer et
al. (1997) came to similar conclusions.  They have proposed that it is the
thermal instability of the HBS that causes episodical rises of the HBS
temperature.

   This having been said, the idea of a hot hydrogen burning origin of the
MgAl anticorrelation in GCRGs also disagrees with the M\,13 magnesium
isotopic analysis of S96 because at $T_6=74$ not only $^{24}$Mg but also
$^{25}$Mg and $^{26}$Mg are quickly destroyed. For example, if we begin with
the summed abundance [$^{25}$Mg+$^{26}$Mg/Fe]\,=\,0 and isotopic ratios
$^{24}$Mg/$^{25}$Mg/$^{26}$Mg\,=\,90/4.5/5.0 (corresponding to the chemical
composition described in Sect.~3.1, and which are also very close to the
ratios observed in the ``unmixed'' M\,13 giant L598 of S96), then after
consumption of 5\% of H we find [$^{25}$Mg+$^{26}$Mg/Fe]\,=\,\mbox{--0.50}
(whereas Shetrone reported values as large as +0.21) and
$^{24}$Mg/$^{25}$Mg/$^{26}$Mg\,=\,92.3/4.7/3.0 (in comparison with S96's
average ratios 56/22/22).  Is this an insuperable problem?  A possible
solution might be to postulate even higher initial abundances of the heavier
Mg isotopes.  What would certainly be most worthwhile is confirmation of the
S96 result, and accurate data for a larger group of objects to more strongly
constrain the situation.

%======================================================================

%==================== Concluding remarks ==============================
\section{Concluding remarks}

   We have shown that the hypothesis of deep mixing in stars ascending the
RGB can explain self-consistently the anomalous abundances of C (as well as
the \cc\ ratios), N, O and Na in \omcen\ giants and the global
anticorrelation of [O/Fe] versus [Na/Fe] most clearly seen in M\,13
giants. There is much other observational evidence of deep mixing in GCRGs,
the most direct being the progressive decline of [C/Fe] with increasing
luminosity on the RGB seen in several globular clusters (see references in
Sect.~1). Similar evidence has been obtained recently by Pilachowski et
al. (1996), who report that red giants in M\,13 become more enriched in Na as
they approach the RGB tip.

   We did not discuss the nature of the mechanism driving the deep mixing.
The most promising candidate seems to be some kind of rotationally induced
instability (baroclinic and/or shear instability etc.). In this case clusters
with especially strong star-to-star abundance variations are expected to
contain a lot of rapidly rotating stars. Indeed, horizontal branch stars in
M\,13 are found to possess unusually fast rotation (Peterson et al. 1995),
and \omcen\ seems to have one of the bluest horizontal branches among the
globular clusters (Whitney et al. 1994) which may be a manifestation of
surface He enrichment during RGB evolution (Rood 1973; Sweigart 1997). Of
course, the latter observational fact does not exclude a primordial origin of
any He enrichment.

   Unfortunately, the deep mixing scenario alone cannot account for the MgAl
anticorrelations in the M\,13 and \omcen\ giants in the absence of additional
{\it ad hoc} assumptions. Possible modifications include a strong but still
undetected low energy resonance in the reaction $^{24}$Mg(p,$\gamma)^{25}$Al,
and episodical increases of the HBS temperature up to the value $T_6\approx
74$ (the standard models predict $T_6\leq 55$) in stars ascending the giant
branch. The first assumption, however, will most probably raise objections by
nuclear physicists, whereas the second disagrees with the results of S96's
magnesium isotopic analysis (Sect.~4.3).  As regards the second possibility,
the work of taking into account feedback from deep mixing on the structure
and evolution of red giants still remains to be done before one can draw any
further conclusions.  Finally, given the importance of Shetrone's results for
the present discussion, we emphasize the importance of their confirmation and
amplification.

   On the other hand, there are very convincing observational arguments in
favour of the primordial scenario. To those mentioned in Sect.~4 one can add
the well-known CN bimodality in 47\,Tuc stars which has been traced down to
the MS turn-off (Briley et al. 1994). Moreover, Briley et al. (1996) have
recently shown that the strong CN (and weak CH) molecular band widths are
accompanied by spectroscopic signatures of increased Na abundance in stars
just below the MS turn-off in 47\,Tuc.  We have tried to determine whether
mixing in a low mass MS star can produce surface C depletion accompanied by N
and Na enhancements without dredging up too much He (which would conflict
with the very narrow CM diagram of 47\,Tuc at the MS turn-off; VandenBerg \&
Smith 1988), but we failed even with the new NeNa-cycle reaction rates of El
Eid \& Champagne (1995).  Hence, the CN bimodality and the Na enhancements in
47\,Tuc stars are most likely of primordial origin and AGB stars might be
responsible for this.

   We have proposed a primordial plus deep mixing scenario in which
intermediate mass AGB stars are considered as the source of a primordial
anticorrelation of $^{24}$Mg versus $^{25}$Mg in GCRGs, with $^{24}$Mg being
depleted in HBB, and $^{25}$Mg being increased in He shell burning during
pulses.  The anticorrelations of [O/Fe] versus [Al/Fe] in M\,13 and \omcen\
giants are then very well reproduced in the deep mixing calculations with Al
synthesized at the expense of $^{25}$Mg. To produce extremely large Al
enhancements, which are observed in some GCRGs, this scenario requires that
the following two {\it necessary} conditions be fulfilled: (1) the star must
have a high initial $^{25}$Mg abundance, and (2) a deep mixing mechanism must
be switched on in the star on the RGB.

   An observational confirmation of our combined scenario would be finding a
red giant which did not show signs of deep mixing (i.e. had large [O/Fe] and
low [Na/Fe] and [Al/Fe]), but at the same time possessed depleted [Mg/Fe] and
increased [$^{25}$Mg+$^{26}$Mg/Fe]. At present there is only one star in
S96's sample, L598, showing no sign of efficient (if at all) mixing.
Unfortunately, with $^{24}$Mg/$^{25}$Mg/$^{26}$Mg\,=\,94/3/3, it does not
possess any trace of pollution from intermediate mass AGB stars. Hence, for
this particular object both of the above conditions are not met. Another test
would be to determine whether the sum $^{25}$Mg+$^{26}$Mg in fact consists of
$^{26}$Mg only, because in our calculations the $^{25}$Mg isotope is
destroyed completely in GCRGs with efficient mixing.

   We comment, finally, on the differences in the abundance patterns between
globular cluster red giants, on the one hand, and the field halo giants, on
the other.  These groups differ in two well documented ways.  First,
anomalies involving C and N are much more prevalent in cluster stars (see
Langer, Suntzeff \& Kraft 1992, and references therein), and, second, it
appears that enhancements of [($^{25}$Mg+$^{26}$Mg)/$^{24}$Mg] exist
preferentially in clusters rather than in the field.  In the context of the
two above-noted conditions, it appears that for field stars one would then
require that neither of them be met.  That is to say, one requires not only
that field stars have not experienced enrichment from AGB stars but also that
they have not experienced deep mixing.  One might argue that economy of
hypothesis suggests it is more palatable to assume, in the context of the
alternatives discussed above, that an agency exists in globular cluster
giants which leads both to deep mixing, on the one hand, and which operates
in conjunction with either the postulated elevated temperatures ($T_6 = 74$)
in the HBS or an accelerated rate for the reaction
$^{24}$Mg(p,$\gamma)^{25}$Al, on the other, and that this does not exist in
the field stars.

   Recently Smith \& Kraft (1996) have considered another combined scenario
where the initial overabundance of $^{25}$Mg in GCRGs is assumed to come from
Ne novae. Our paper extends the list of primordial plus deep mixing
scenarios. What will be next?
 
%======================================================================

%==================== Acknowledgements ================================
\begin{acknowledgements}

   P.A.D. wishes to express his gratitude for the warm hospitality of the
staff of the Max-Planck-Institut f\"{u}r Astrophysik where the s-process
nucleosynthesis code was prepared and appreciates the very favourable working
atmosphere at the Mount Stromlo \& Siding Spring Observatories where this
study was carried out.  P.A.D., G.S.Da C. and J.E.N. gratefully acknowledge
financial support from the Department of Industry, Science \& Tourism of the
Australian Government, for a visit of P.A.D. to Australia, during which this
work was performed.  Our special thanks go to A.I.Boothroyd, J.C.Lattanzio,
K.Takahashi, C.Tout and S.E.Woosley.

\end{acknowledgements}

%==================== Bibliography ====================================
{}

\end{document}